\documentclass[11pt]{article}

\usepackage{cite}
\usepackage{xspace}

\newtheorem{theorem}{Theorem}
\newtheorem{lemma}[theorem]{Lemma}
\newtheorem{corollary}[theorem]{Corollary}
\newenvironment{proof}{\noindent\par{\bf Proof:
}}{\nopagebreak\rule{1 ex}{0.8 em}\medskip}

\newcommand{\algname}{{\sc{lean-consensus}}\xspace}

\newcommand{\newloglike}[2]{\newcommand{#1}{\mathop{\rm #2}\nolimits}}
\newloglike{\E}{E}

\newcommand{\remove}[1]{}

\newcommand{\floor}[1]{\left\lfloor{#1}\right\rfloor}

\newcommand{\rmax}{r_{\max}}

\begin{document}

\title{Fast Deterministic Consensus in a Noisy Environment}

\author{James Aspnes\thanks{
Yale University, Department~of
Computer Science, 51 Prospect Street/P.O. Box 208285, New
Haven CT 06520-8285.
Email: \texttt{aspnes@cs.yale.edu}.
This work was supported in part by NSF grants CCR-9820888 and
CCR-0098078.}}

\maketitle

\begin{abstract}
It is well known that the consensus problem cannot be solved
deterministically in an asynchronous environment, but that
randomized solutions are possible.
We propose a new model, called \emph{noisy scheduling}, in which an
adversarial schedule is perturbed randomly, and show
that in this model randomness in the environment can substitute for
randomness in the algorithm.
In particular, we show that
a simplified, \emph{deterministic} version of 
Chandra's wait-free shared-memory consensus algorithm
(PODC, 1996, pp.~166--175)
solves consensus 
in time at most logarithmic in the number of \emph{active} processes.
The proof of termination
is based on showing that a race between independent
delayed renewal processes produces a winner quickly.
In addition, we show that the protocol finishes in constant time using
quantum and priority-based scheduling on a uniprocessor, suggesting
that it is robust against the choice of model over a wide range.
\end{abstract}

\section{Introduction}

Perhaps the single most dramatic result in the theory of distributed
computing is Fischer, Lynch, and Paterson's proof of the impossibility
of deterministic consensus in an asynchronous environment with
failures \cite{FischerLP85}.  This result and its extensions 
\cite{LouiA1987,DolevDS87}
show that the consensus
problem, in which a group of processes must collectively agree on a
bit, cannot be solved deterministically in an asynchronous message-passing
or shared-memory model if an unrestricted adversary
controls scheduling.  
Solutions to the shared-memory version of this fundamental
problem have
thus taken the approach of restricting the adversary, either by
allowing randomization that limits the adversary's knowledge
\cite{ChorIL1994,Abrahamson1988,AttiyaDS1989,AspnesH90,SaksSW91,Aspnes93,BrachaR1991,Chandra1996,AumannB1996,Aumann1997} 
or by
imposing timing constraints that limit the adversary's control
\cite{DolevDS87,DworkLS88,AlurAT1997}.  
As a corollary to granting less power to the
adversary, these solutions often involve granting more power to the
algorithm, in the form of the ability to obtain random
bits or explicitly delay steps.
By using these additional powers an algorithm can escape the FLP bound
and reach agreement.

These additional powers come at a cost.  Randomization alone is
not powerful enough to allow sublinear consensus protocols
\cite{Aspnes1998}, so
efficient randomized solutions have required additional constraints on
the ability of the adversary to observe the arguments to operations and
the contents of unread memory locations
\cite{Chandra1996,AumannB1996,Aumann1997}.
These algorithms carefully manage common pools of unread
random bits for future use, a clever but odd-looking practice that
is justified primarily by the specific details of the model.  The delay-based
algorithm of \cite{AlurAT1997} is less convoluted, but still depends on
using explicit delays that at the minimum require 
that a process has the power to
invoke them and at worst may add unnecessary delay when
few processes participate.

As an alternative to designing an algorithm specifically to exploit
the weaknesses of a particular adversary model, we consider the
approach of using a simple algorithm that guarantees agreement but
relies on good luck to terminate.
Our \algname{} algorithm, described in Section
\ref{section-algorithm}, is obtained by removing all of the
randomized parts of a similar algorithm due to Chandra
\cite{Chandra1996}.  
The essential idea (which is the core of many consensus protocols in the
literature) is to stage a race between those processes that
prefer $0$ and those that prefer $1$, with the rule that
if a slow process sees that
faster processes are all in agreement it adopts their common
preference.
The race is implemented using two arrays of atomic read/write bits.
The algorithm terminates when the fastest processes are all in
agreement and can decide on their preferred value
safely, knowing that other processes will
adopt the same preference before they catch up.
As shown in Section~\ref{section-agreement}, this mechanism is enough
to ensure that if any one process
decides then all other processes soon decide on the same value,
no matter how the adversary arranges the schedule.

In effect, the race framework allows the processes to detect agreement
once it occurs.
But unlike other consensus algorithms, \algname{} makes no attempt to
cajole the processes into reaching agreement--- it relies
entirely on the hope that some process eventually pulls ahead of the
others.  In order to dash this hope, the adversary must exercise enough
control to ensure that the fastest processes run in lockstep.  We
believe that in many natural system models it will be difficult for the
adversary to exercise this much control.

One such model is what we call the
\emph{noisy scheduling} model, 
described in Section
\ref{section-noisy}.  In this model, the
adversary proposes a schedule 
that specifies the order in which read and write operations occur,
but this schedule is perturbed by random
noise drawn from some arbitrary non-constant distribution.
This noise corresponds to random factors in a system that might not
be
strongly correlated with the algorithm's behavior, such as network
delays, clock skew, or bus or memory contention.

We show in Section~\ref{section-noisy-termination} 
that, in the noisy scheduling model, \algname{} terminates with expected
$\Theta(\log n)$
work per process, where $n$ is the number of active
processes.
This result is distribution-independent, in the sense that the
algorithm's asymptotic performance does not depend on the noise
distribution in the model (though the constant factor does), and it 
holds even if processes are subject to random halting failures.
Because the algorithm's performance depends only on the
number of processes actually executing the protocol and not on 
the total number of processes in the system, 
it is \emph{adaptive} in the sense of \cite{AttiyaF1998},
which implies
it is \emph{fast} in the
sense of 
\cite{Lamport87,AfekDT95}.
Thus it is well-suited to situations where only one or a few processes
attempt to run the algorithm at the same time.

Our noisy scheduling model
is similar to the model used by Gafni and Mitzenmacher
\cite{GafniM1999}
in their analysis of mutual exclusion protocols with random timing, but is
extended to include constant delays inserted by the adversary in
addition to random delays.
Another source of inspiration is Koutsoupias and Papadimitriou's
\emph{diffuse adversary}
\cite{KoutsoupiasP94}, 
which chooses a distribution over executions in which no branch at any
decision point can occur with probability more than some fixed
$\epsilon$.
Our model is not the first in which an adversary chooses
parameters for a stochastic process that then controls scheduling; a
sophisticated model of this type, based on asynchronous PRAMs, has been
proposed by Cole and Zajicek \cite{ColeZ95}.

To give support to our intuition that many possible restrictions
on the adversary make \algname{} work, we also consider what happens
with a hybrid quantum and priority-based scheduler on a uniprocessor,
following the approach of \cite{AndersonM1999}.  (The details of this
model, which subsumes both quantum scheduling and priority-based
scheduling, are sketched in Section \ref{section-hybrid}.)  We show in
Section \ref{section-hybrid-termination} that \algname{} terminates in
$O(1)$ steps in the hybrid-scheduling
model, as long as the quantum is at least 8.
The restriction to a uniprocessor is
necessary because \cite{AndersonM1999} shows that no
deterministic algorithm can solve consensus with multiple processors,
even with hybrid scheduling,
without using stronger primitives than atomic read/write registers.

Our basic consensus algorithm requires infinitely long arrays.
Obviously this is
undesirable in a real system.  In order to bound the required space,
we adopt a technique from \cite{Chandra1996} and
cut off the algorithm after consuming $O(\log^2 n)$ bits
of space,
using the preference each undecided process has at that point as input to a
more expensive, 
bounded-memory consensus algorithm satisfying the validity
property.\footnote{An early example of this approach 
is found in
the bounded-rounds randomized Byzantine agreement
protocol of Goldreich and Petrank~\cite{GoldreichP1990}, which
switches from a randomized to a deterministic protocol if the
randomized protocol does not terminate quickly enough.}
Since the more expensive algorithm is only run with low probability,
its higher costs do not increase the expected time for the algorithm as
a whole by more than a small constant factor.
Details are given in Section~\ref{section-bounded-memory}.

Section~\ref{section-simulations} 
describes some simulation results
that show that the constant factors in the noisy scheduling
analysis are in fact quite small for plausible noise distributions,
suggesting that the good theoretical performance of \algname{} might actually
translate into fast execution in a real system.

In Section \ref{section-extensions}, we suggest a number of directions
in which the current work could be extended, including extensions to
the noisy scheduling model.
One interesting possibility is the inclusion of adaptive crash
failures.  We
argue briefly that 
because \algname{} recovers quickly from such failures,
it terminates in at most $O(f \log n)$ work per process even if up to
$f$ processes fail.  However, there remains an interesting open
question whether noisy scheduling is enough to get $O(\log n)$
performance even with $\Theta(n)$ crash failures.

\section{The Consensus Problem}

In the \emph{binary consensus problem}, a group of $n$
processes, possibly subject to halting failures,
must agree on a bit.\footnote{Some authors 
consider the stronger problem of \emph{id consensus},
in which the decision value is the id of
some active process.  In many cases, id consensus can be
solved in a natural way using a $(\lg n)$-depth tree of binary
consensus protocols; examples of this approach can be found in
\cite{Chandra1996,Aumann1997}.}
A \emph{consensus protocol} is a distributed algorithm in which each
non-faulty
process starts with an input bit and eventually terminates by deciding
on an output bit.  
It must satisfy
the following three conditions with probability 1:

\begin{itemize}
\item \emph{Agreement.} All non-faulty processes decide on the same bit.
\item \emph{Termination.} All non-faulty processes finish the protocol in
a finite number of steps.
\item \emph{Validity.} If all processes start with the same input bit,
all non-faulty processes decide on that bit.\footnote{
Some definitions of consensus replace the validity condition with a
weaker \emph{non-triviality} condition that says that there must exist
executions in which different decision values occur.  
}
\end{itemize}

\section{Model}

We assume a shared-memory system consisting of an unbounded number of
processes that communicate only through shared atomic read/write
registers.
We use the usual interleaving model, in which operations are assumed to
occur in a sequence $\pi_1,\pi_2,\ldots$, and in which each read
operation returns the value of the last previous write to the same
location.  The order in which operations occur is determined by a
stochastic process that is partially under the control of an
adversary (Section \ref{section-noisy}), or directly by the adversary
subject to certain regularity constraints (Section
\ref{section-hybrid}).

\subsection{Noisy Scheduling}
\label{section-noisy}

In the \emph{noisy scheduling} model, we assume that the adversary
specifies when operations occur (subject to an upper bound on the time
between successive operations by the same process), but that this
specification is perturbed by random noise.

Formally, the adversary chooses:

\begin{enumerate}
\item An arbitrary starting time $\Delta_{i0}$ for each process $p_i$,

\item A non-negative 
delay $\Delta_{ij}$ between process $p_i$'s $(j-1)$-th and $j$-th
operations, bounded by some fixed constant $M$, and

\item A fixed common distribution $F_{\pi}$ of the random delay 
added to
each type of operation
$\pi$ (e.g., read or write).
If process $p_i$'s $j$-th operation is of type $\pi$,
it suffers an additional delay $X_{ij}$ whose distribution is $F_\pi$.
There is no restriction on the choice of 
the $F_\pi$,
except that they must not be concentrated on a point and must produce
only non-negative values
$X_{ij}$.\footnote{In fact, the $F_\pi$ distributions can be quite
bizarre; it is not required, for example, that the $X_{ij}$ have finite
expectation.}
\end{enumerate}

The time of process $p_i$'s $j$-th operation is given by
\begin{displaymath}
S_{ij} = 
\Delta_{i0} + \sum_{k = 1}^{j} \left(\Delta_{ik} + X_{ik}\right).
\end{displaymath}

Since we are using interleaving semantics, the effect of executing two
operations at exactly the same time is not well-defined.  To avoid
ill-defined executions, we
impose the additional technical constraint on the adversary's choices that
the probability that any two operations occur
simultaneously must be zero.  This is automatic if, for example, the
noise distributions $F_\pi$ are continuous.  Alternatively, it can
be arranged by dithering the starting times of each process by some
small epsilon.  This technical constraint does not qualitatively change
our results.

Below we discuss the unfairness 
of noisy scheduling and extensions to
allow random failures.

\subsubsection{Unfairness}

The upper bound on the $\Delta_{ij}$
and the common distribution on the $X_{ij}$ might suggest that the
noisy scheduling model produces fair schedules.  This is not entirely
true for sufficiently pathological distributions.
\begin{theorem}
There exists a choice of $F_\pi$ and $\Delta_{ij}$ such that for any
distinct processes $p_i$ and $p_{i'}$, and any operation $j$,
the expected number of operations $p_{i'}$ completes between
$p_i$'s $j$-th and $(j+1)$-th operations is infinite.
\end{theorem}
\begin{proof}
Set each $F_\pi$ so that
$X_{ij}$ takes on the value $2^{k^2}$ with probability
$2^{-k}$ for $k = 1, 2, \ldots$.  For simplicity, let us suppose 
that $\Delta_{ij}=0$ for $j > 0$.  We will also assume that $A$ and $B$ 
execute no operations before time $0$.

Let $X$ be the number of operations completed by $p_{i'}$ between
$S_{ij}$ and $S_{i,j+1}$.  We will show that the expectation of $X$ is
infinite conditioned on the value of
$t = \lceil S_{ij} \rceil$ (the ceiling is so that we have countably
many cases).

The idea is this: for each $k$ we have probability $2^{-k}$ that
$S_{i,j+1} \ge X_{i,j+1} = 2^{k^2}$.  
Condition on this event occurring for some particular $k$ and
consider how many operations $p_{i'}$ must execute to reach time
$2^{k^2}$.  Either (a) one of these operations
takes time $2^{k^2}$ or more (with
probability $2^{-k+1}$ per operation); or 
(b) a total of at least $2^{2k-1}$ faster
operations, each of which
takes at most $2^{(k-1)^2}$ time, must occur.
If we wait only for event (a), we expect to see $2^{k-1}$ operations;
to get the actual expected number, we must subtract off the expected
number of operations until (a) occurs after (b) occurs ($2^{k-1}$
again) multiplied by the probability that (b) occurs.  This latter
probability is at most 
$(1-\frac{1}{2^{k-1}})^{2k-1}$, which goes to $e^{-2}$ in the limit as
$k$ grows; it follows that $p_{i'}$ executes $\Omega(2^k)$ operations
on average before time $2^{k^2}$.  Of these, at most $t/2$ can occur
before time $S_{ij}$, so if $k \gg \lg t$, we have $\Omega(2^k)$
operations on average between $t$ and $2^{k^2}$, and thus also between
$S_{ij}$ and $S_{i,j+1}$, since $S_{ij} \le t < 2^{k^2} \le S_{i,j+1}$.

To get the full result, we must remove two layers of conditioning.
First compute the expectation conditioned only on $t$ by
summing $2^{-k} \Omega(2^k)$ for each of the infinitely many
sufficiently large $k$.  It is
not difficult to see that this sum diverges and the expectation is
infinite.  Summing over all values of $t$ doesn't make it any less
infinite, and we are done.
\end{proof}

\subsubsection{Failures}

We can extend the noisy scheduling model to allow halting failures.
For each $i$ and each $j > 0$ let $H_{ij} = \infty$ if process $p_i$
halts before its $j$-th operation and $0$ otherwise.  Define
\begin{displaymath}
S'_{ij} = 
\Delta_{i0} + \sum_{k = 1}^{j} \left(\Delta_{ik} + X_{ik} +
H_{ik}\right),
\end{displaymath}
with the usual convention for the extended real line 
that $x+\infty = \infty+x = \infty$ for any finite $x$.
If $S'_{ij} = \infty$, $p_i$'s $j$-th operation does not occur.

We do not include failures in the noise distributions $F_\pi$ because
these distributions do not depend on $n$, and a constant probability of
failure would mean that all processes die after $O(\log n)$ steps.
Instead, we assume that failures occur
independently with probability $h(n)$ per operation, where $h$ is some
function chosen by the adversary.  The effect of stronger failure
models is discussed in Section \ref{section-extensions}.

\subsection{Quantum and Priority-Based Scheduling}
\label{section-hybrid}

Our intuition is that \algname{} should perform well in any setting
that prevents lockstep executions.  One such setting is the
hybrid-scheduled
uniprocessor model of \cite{AndersonM1999}, which combines the
priority-based scheduling model of \cite{RamamurthyMA1996}
with the quantum-based
scheduling model of \cite{AndersonJO1998}.  In this model, processes are
assumed to be time-sharing a uniprocessor under the control of a
pre-emptive scheduler.  Each process has a priority, and a process may
be pre-empted at any time by a process of higher priority.  A process
may only be pre-empted by a process of the same priority if it has
exhausted its \emph{quantum}, a minimum number of operations it must
complete between the time it wakes up and the time at which it becomes
vulnerable to pre-emption.  There is no requirement that a process
start the protocol at the beginning of a quantum; it may have used up
some or all of its quantum performing other work before starting the
protocol.  We do not consider failures in the hybrid-scheduling model; instead,
a process may be arbitrarily delayed subject to the constraints on the
scheduler.

\section{The \algname{} Algorithm}
\label{section-algorithm}

In this section, we describe the \algname{} algorithm.
The algorithm is very simple, because we are relying on 
randomness in the environment
to guarantee termination and thus the algorithm itself must
only guarantee correctness and provide the opportunity for the
underlying system to quickly jostle it into a decision state.
Structurally, it is essentially identical
to the 
multi-writer register consensus protocol of Chandra
\cite{Chandra1996}
with the shared coins removed,
leaving only the implementation from multi-writer bits
of the ``racing counters'' technique that has
been used in many shared-memory consensus protocols.
It also bears some similarities
to the Time-Adaptive Consensus algorithm of Alur et
al.\cite{AlurAT1997} with the delays removed.

At each step of the algorithm, each process \emph{prefers} either 0 or 1 as
its decision value.  The conflict between the 0-preferring processes and
the 1-preferring processes is settled by a race implemented using two
arrays $a_0$ and $a_1$ of atomic read/write bits, each initialized to
zero.  Each process
carries out a sequence of rounds, each consisting of a fixed sequence
of operations.  During round $r$, a process that prefers $b$ marks
location $a_b[r]$ with a one
and looks to see if either (a) it has fallen behind
its rivals who prefer $(1-b)$, in which case it abandons its former
preference and joins the winning team, or (b) it and its fellows have
sped far enough
ahead of any rival processes that they can safely decide $b$ knowing
that those rivals will give up and join the $b$ team before they catch
up.
The algorithm finishes fastest when the pack of processes disperses
quickly, so that a clear winner emerges as early as possible.

Let us look more closely at the details of the algorithm.
A process with input $b$ sets its preference $p$ to $b$ 
and its round number $r$ to $1$.
(We say that a process is \emph{at
round $r$} if its round number is set to $r$; processes thus start at round
$1$.)
It then repeatedly executes the following
sequence of steps.  To simplify the description of the algorithm, we
assume that while $a_0$ and $a_1$ are initialized to zeroes,
they are prefixed with (effectively read-only)
locations $a_0[0]$ and
$a_1[0]$,
both set to $1$.

\begin{enumerate}
\item Read $a_0[r]$ and $a_1[r]$.  If for some $b$,
$a_b[r]$ is $1$ and $a_{1-b}[r]$
is $0$, set $p$ to $b$.
\item Write $1$ to $a_p[r]$.
\item Read $a_{1-p}[r-1]$.  If this value is $0$, decide $p$ and exit.
\item Otherwise, set $r$ to $r+1$ and repeat.
\end{enumerate}

Note that in each round the process carries out exactly four
operations in the same sequence: two reads, a write, and another read.
It is tempting to optimize the algorithm by eliminating the write when
it is already evident from the previous step that $a_p[r]$ is set or
eliminating the last read when it can be deduced from the value of
$a_{1-p}[r]$ that $a_{1-p}[r-1]$ is set.  
However, this optimization reduces the work done by slow processes
(whom we'd like to have fall still further behind) while maintaining the same
per-round cost for fast processes (whom we'd like to have pull 
ahead).  So we must paradoxically carry out operations that 
might appear to be superfluous in order to minimize the actual
total cost.

\section{Agreement and Validity}
\label{section-agreement}

If we ignore the termination requirement, the correctness of the
algorithm does not depend on the behavior of the scheduler.  The
following two lemmas show that the validity and agreement properties
hold whenever the algorithm terminates.  The proofs are very similar
in spirit to those of Lemmas 1-4 in \cite{Chandra1996}.

\begin{lemma}
\label{lemma-converging-preferences}
No process sets $a_{b}[r]$ unless (a) $r = 1$ and $b$ is an input
value, or (b) $r > 1$ and $a_{b}[r-1]$ has already been set.
\end{lemma}
\begin{proof}
Consider the first process $P$ that sets $a_{b}[r]$.  
Then $P$ does not read $1$ from $a_{b}[r]$ at round $r$ and does not
change its preference during round $r$.  If $r=1$, $P$'s preference
equals its input, establishing case (a); if $r > 1$, $P$ must have set
$a_{b}[r-1]$ at round $r-1$, establishing case (b).
\end{proof}

\begin{lemma}
\label{lemma-validity}
If every process starts with the same input bit $b$, every process
decides $b$ after executing 8 operations.
\end{lemma}
\begin{proof}
From Lemma~\ref{lemma-converging-preferences},
if no process has input $1-b$, no process ever sets $a_{1-b}[1]$.  It
follows that every process sees a zero in $a_{1-b}[1]$ at round 2 and
decides $b$.
\end{proof}

\begin{lemma}
\label{lemma-agreement}
If some process decides $b$ at round $r$, 
then 
(a) no process ever writes $a_{1-b}[r]$,
and
(b) every process decides $b$ at
or before round $r+1$.
\end{lemma}
\begin{proof}
Let $P$ decide $b$ at round $r$.
We will show that this implies that no process ever sets $a_{1-b}[r]$.  

Suppose some process sets $a_{1-b}[r]$; let $Q$ be the first such
process.  Because $Q$ is the first process to set $a_{1-b}[r]$, it must read a
$0$ from $a_{1-b}[r]$ at the start of round $r$.  Thus $Q$ can only set
$a_{1-b}[r]$ if it already prefers $1-b$ at the start of round
$r$, implying that it set $a_{1-b}[r-1]$ during round $r-1$; 
and if it reads a $0$ from $a_{b}[r]$ at the start of round $r$,
preventing it from changing its preference after seeing a $0$ in
$a_{1-b}[r]$.
But $Q$'s read of $a_{b}[r]$ occurs after $Q$'s write to
$a_{1-b}[r-1]$, which occurs after $P$'s read of $a_{1-b}[r-1]$ at
round $r$ (because $P$ reads $0$), 
which in turn occurs after $P$'s write to $a_b[r]$.  Thus
$Q$ reads $1$ from $a_b[r]$, and changes its preference to $b$ at
round $r$.  This contradicts our
assumption that $Q$ is the first to set $a_{1-b}[r]$. 
It follows that if any process decides $b$ in round $r$, no process sets
$a_{1-b}[r]$.

Since no process sets $a_{1-b}[r]$, any process that reaches round
$r+1$ must set $a_{b}[r+1]$ (by Lemma~\ref{lemma-converging-preferences}),
and will decide $b$ after reading $0$ from $a_{1-b}[r]$.
Thus no process runs past round $r+1$ without deciding $b$.

To show agreement in earlier rounds, let $P'$ decide $b'$ at round $r'
\le r$.  By the preceding argument, if $P'$ decides $b'$ at round
$r'$, then no process sets $a_{1-b'}[r']$ and thus (by
Lemma~\ref{lemma-converging-preferences} again) no process sets
$a_{1-b'}[r]$.  But since $P$ sets $a_{b}[r]$, we must have $b'=b$.
\end{proof}

\section{Termination with Noisy Scheduling}
\label{section-noisy-termination}

In this section, we show that \algname{} terminates in $\Theta(\log n)$
rounds with noisy scheduling and random failures.  (This analysis includes the
core model without random failures as well, since the adversary can
always choose $h(n)=0$.)
We show that
either all processes die (in which case we treat the algorithm as
terminating in the last round in which some process takes a step), or some
group of processes with a common preference eventually gets two
rounds ahead of the other processes.  To avoid analyzing the
details of how processes shift preferences, we will show the even
stronger result that unless all processes die,
a \emph{single} process eventually gets two rounds
ahead of the other processes.  

To simplify the argument, we abstract away from the individual
sequence of operations in each round and look only at the times at
which rounds are completed.  We can thus assume that the adversary
provides a single noise distribution $F$ (corresponding to the
distribution of the sum of the delays on three reads and one write)
and that the values $\Delta_{ij}$, $X_{ij}$, and $H_{ij}$ 
provide the delay not
on the $j$-th \emph{operation} but on the $j$-th \emph{round}.
Since this abstraction merely involves summing together the underlying
variables on operations, it does not reduce the adversary's control
over the protocol.  We will scale $M$ appropriately so that it is
still the case that $0 \le \Delta_{ij} \le M$ when $j > 0$.

Using this approach, the increment $\Delta_{ij} + X_{ij} + H_{ij}$ 
is the time taken for
process $i$ to move from the end of round $j-1$ to the end
round $j$.  The
constant $\Delta_{i0}$ represents the process's starting time, and 
$S'_{ir} = \Delta_{i0} + \sum_{j=1}^{r} \left(\Delta_{ij} +
X_{ij} + H_{ij}\right)$
gives the time at which the process finishes round $r$.  A process $i$ wins
the race with a lead of $c$ rounds
at round $r+c$ if it finishes round $r+c$ before any other
process finishes round $r$, i.e., if $S'_{i,r+c} \le S'_{i',r}$ for all
$i' \ne i$.

We would like to show a bound on how the expected round at which some
process wins by $c$ scales as a function of the number of processes $n$,
keeping $c$, $M$, and $F$ fixed.  
This bound is given in Corollary \ref{corollary-race} below.
We will assume that $h(n) = o(1)$, as otherwise all processes die after
$O(\log n)$ rounds on average.
The proof proceeds in two steps: first we show that for \emph{any} $r$
which some process finishes with at least constant probability, there
exists a critical time $t$ that gives at least a constant probability
that $S'_{ir} \le t$ for exactly one $i$.  We then show that if $r$ is
large enough, $\Pr[S'_{i,r+c} \le t | S'_{ir} \le t]$ is also at least a
constant.  It then follows that the probability that $S'_{i,r+c} \le t$
while
$S'_{i'r} > t$ for any $i' \ne i$ is at least the product of these two
constants and the constant probability that $p_i$ is not killed
between rounds $r$ and $r+c$.  
Thus after a constant number of phases each consisting
of $r+c$ rounds we expect some process to win.

\subsection{Existence of a winner}

In this section, we build up the tools needed to show that for each
round there exists a fixed time at which there is likely to be a
unique winner.

\begin{lemma}
\label{lemma:one-vs-zero}
Let $A_1, \ldots, A_n$ be independent events.
If the probability that no $A_i$ occurs is $x$,
where $x$ is not zero,
then the probability that exactly one $A_i$ occurs is
at least $-x \ln x$.
\end{lemma}
\begin{proof}
Let $q_i$ be the probability that $A_i$ does not occur.
The probability $x$ that no $A_i$ occurs is the product of the $q_i$.
Since $x$ is nonzero, each $q_i$ must also be nonzero.
The probability that exactly one $A_i$ occurs is given by
\begin{eqnarray}
\left(\prod_{i=1}^{n} q_i\right) \sum_{i=1}^{n} \frac{1-q_i}{q_i}
&=&
x \sum_{i=1}^{n} \left(\frac{1}{q_i} - 1 \right)
\nonumber
\\
&=&
x \left(-n + \sum_{i=1}^{n} \frac{1}{q_i}\right).
\label{eq:one-vs-zero}
\end{eqnarray}

Let $G$ be the geometric mean of the $q_i$ and let $H$ be their
harmonic mean.  By the theorem of the means, $G > H$.
Observe that $G = x^{1/n}$ and
\begin{displaymath}
\sum_{i=1}^{n} \frac{1}{q_i} = n/H
> n/G = n x^{-1/n} = n \exp\left(-\frac{\ln x}{n}\right)
\ge n \left(1 - \frac{\ln x}{n}\right)
= n - \ln x.
\end{displaymath}
Plugging this inequality into (\ref{eq:one-vs-zero}) gives the result.
\end{proof}

Suppose $X_1, \ldots, X_n$ are random times.  The following lemma
shows that under certain conditions there exists a constant time
$t_0$,
such that, with constant probability, at most one of the $X_i$ is less
than $t_0$:

\begin{lemma}
\label{lemma-winner}
Let $X_1, \ldots, X_n$ be
independent random variables such that for
all finite values $t$ and all distinct $i, j$, the probability that
$X_i = X_j = t$ is zero.
Then either $\Pr[\forall i X_i = \infty]$ is greater than $e^{-1}$ or
there exists $t_0$ such that the
probability that exactly one of the $X_i$
is less than or equal to $t_0$
is at least $1/5$.
\end{lemma}
\begin{proof}
For each $t$, let $q_i(t)$ be the probability that $X_i$ is not
less than or equal to $t$.
Let $q(t) = \prod_{i=1}^{n} q_i(t)$
be the probability that none of the $X_i$
are
less than or equal to $t$.
Note that each $q_i(t)$ is
a decreasing right-continuous left-limited function with
$\lim_{t \rightarrow -\infty} q_i(t) = 1$
and
$\lim_{t \rightarrow \infty} q_i(t) = \Pr[X_i = \infty]$.
Similarly, $q(t) = \prod_{i} q_i(t)$ is right-continuous,
left-limited, and has
$\lim_{t \rightarrow -\infty} q(t) = 1$
and
$\lim_{t \rightarrow \infty} q(t) = \Pr[\forall i X_i = \infty]$.

Suppose that this latter quantity is less than or equal to $e^{-1}$.
(If not, the first case of the lemma holds.)
Then for some finite $t$, $q(t) \le e^{-1}$.
Let $t_0$ be the least such $t$.

Now suppose $q(t_0) \ge e^{-2}$.
Then, by Lemma \ref{lemma:one-vs-zero},
the probability that exactly one $X_i$ is less than or equal to $t_0$
is at least $2 e^{-2} \approx 0.27\ldots$.

Otherwise, we have $q(t_0) < e^{-2}$ but
$q(t_0-) = \lim_{t \rightarrow t_0-} q(t) > e^{-1}$.
(We are using the usual convention
that $f(x-)$ denotes the left limit
of $f$ at $x$.)
This discontinuity must correspond to a discontinuity
in $q_i$ for some $i$.
At most one $q_i$ has a discontinuity at $t_0$,
by the assumption that the probability that distinct $X_i$, $X_j$
both equal $t_0$ is zero.
Hence, for all $j \ne i$ we have
$q_j(t_0-) = q_j(t_0)$
and thus
$q_i(t_0-)/q_i(t_0) = q(t_0-)/q(t_0) \le e^{-1}$.

Since $q_i(t_0-) \le 1$, it follows immediately that
$q_i(t_0) \le e^{-1}$
and thus the probability that $X_i$
is less than or equal to $t_0$
is at least $1 - e^{-1}$.
Now the probability that no other $X_j$
is less than or equal to $t_0$
is at least $q(t_0)/q_i(t_0) \ge q(t_0-) > e^{-1}$.
Since the variables are independent, the probability that
only $X_i$
is less than or equal to $t_0$
is thus at least $(1-e^{-1})e^{-1} \approx 0.23\ldots$.
\end{proof}

\subsection{Size of the lead}

In this section, we show that if enough rounds have passed, a
process that is likely to be ahead of the others is in fact likely to
be several rounds ahead.  The proof is somewhat complicated by the
lack of restrictions on the noise distribution, but the following
lemma shows how the Strong Law of Large Numbers can be used to smooth
the noise terms out a bit.

\begin{lemma}
\label{lemma-gap}
Let $X_1, X_2, \ldots$ be finite non-negative independent identically
distributed random variables whose common distribution is
not concentrated on a point.  Define $S_n =
\sum_{i=1}^{n} X_i$.  For any $c$, there exist
$n, t$ such that $\Pr[S_n < t] < \frac{1}{2}$
but $\Pr[S_n < t-c] > 0$.
\end{lemma}
\begin{proof}
Let us first consider the case where $X_i$ has a finite expectation
$m$.  Then the Strong Law of Large Numbers says that 
$S_n/n$ converges to $m$ in the limit with probability $1$.  
So for any $\epsilon > 0$, the probability that $S_n$ is less than
$m-\epsilon$ goes to zero and thus drops below $1/2$ for all $n$
greater than some $n_0$.

Let $t_n = n(m-\epsilon)$.  As long as $n > n_0$, we have 
$\Pr[S_n < t] < \frac{1}{2}$.
Now suppose that $\Pr[S_n < t_n-c] = 0$ whenever $n > n_0$.  Since the
$X_i$ are independent, this event can only occur if for each $X_i$,
$X_i < \frac{t_n - c}{n} = m-\epsilon-\frac{c}{n}$ with probability $0$.
Taking the union of countably many 
such bad events for each rational $\epsilon$ and
each $n > n_0$ shows that the event $X_i < m$,
also has probability $0$.  It follows that $X_i \ge E[X_i]$ almost
surely and thus the distribution of $X_i$ is concentrated on
$E[X_i]$, a contradiction.

If $X_i$ does not have a finite expectation, then $S_n/n$ grows without
bound with probability $1$ (see the corollary to Theorem 22.1 in
\cite{Billingsley}).  So for any $x$, there exists $n_0$, such that
$\Pr[S_n/n < x] < \frac{1}{2}$ for $n > n_0$.  We repeat the above
analysis for $t = nx$; if $\Pr[S_n < t-c] = 0$ for all such $t$, we get
$X_i \ge x - \frac{c}{n}$ almost surely, implying $X_i$ exceeds any
finite bound $x$.  Again, a contradiction.
\end{proof}

Once the noise terms have been smoothed, it is not hard to show that
they eventually accumulate enough to push a winner ahead:

\begin{lemma}
\label{lemma-gap-relative}
Fix $c > 0$.  
Let $X_1, X_2, \ldots$ be finite independent identically distributed random
variables such that there exists a threshold $t_0$ for which
$\Pr[X < t_0] < \frac{1}{2}$ but
$\Pr[X < t_0-c] = \delta_0 > 0$.
Define $S_n = \sum_{i=1}^{n} X_i$.

Then for any $\epsilon > 0$, there exists an $n =
O(\log(1/\epsilon))$,
such that for any $t$,
$\Pr[S_n < t] > \epsilon$
implies
$\Pr[S_n < t - c | S_n < t] > \frac{1}{7}\delta_0$.
\end{lemma}
\begin{proof}
Set $n = 8(\ln(1/\epsilon)+1)$.  Each $X_i$ has probability at most
$1/2$ of being less than $t_0$, so a simple application of Chernoff
bounds shows that the probability that 3/4 or more of the $X_i$ are
less than $t_0$ is at most
$e^{-n/8} = \epsilon/e$.

We will use this fact to argue that even when conditioning on $S_n <
t$, there is nearly one chance in four that $X_n$ in particular is
greater than $t_0$.  In this case, $S_{n-1}$ is less than $t-t_0$ and
we can use independence to replace $X_n$ with a new value less than
$t_0 - c$, giving a sum $S_n$ less than $t-c$, all without reducing the
probability by much.

Formally, we have the following sequence of inequalities, each of
which is implied by the previous one.  Let 
$\Pr[S_n < t] = p$ and suppose $p > \epsilon$.  Then we have:

\begin{eqnarray*}
\Pr[S_n < t] &=& p \\
\Pr[S_n < t \wedge 
\mbox{at least $\frac{1}{4}$ of $X_i$ are
greater than $t_0$}]
&>& p-\epsilon/e \\
\Pr[S_n < t \wedge X_n > t_0] &>& \frac{1}{4}(p-\epsilon/e) \\
\Pr[S_{n-1} < t - t_0] &>& \frac{1}{4}(p-\epsilon/e) \\
\Pr[S_{n-1} < t - t_0 \wedge X_n < t_0-c] 
&>&
\frac{1}{4}(p-\epsilon/e)\delta_0 \\
\Pr[S_n < t - c]
&>&
\frac{1}{4}(p-\epsilon/e)\delta_0 \\
\Pr[S_n < t - c | S_n < t]
&>&
\frac{1}{4}(p-\epsilon/e)\delta_0/p
\end{eqnarray*}

Since $p > \epsilon$, this last quantity is at least
$\frac{1}{4}(1-1/e)\delta_0$, which is in turn greater than
$\frac{1}{7}\delta_0$.
\end{proof}

We can now combine Lemmas \ref{lemma-gap} and
\ref{lemma-gap-relative} into the following:

\begin{lemma}
\label{lemma-gap-combined}
Let $X_1, X_2, \ldots$ be finite
non-negative independent identically distributed
random variables whose common distribution is not
concentrated on a point.  Define $S_n = \sum_{i=1}^{n} X_i$.
Fix $c > 0$.  Then there is a constant $\delta$,
such that for any $\epsilon > 0$, there exists $n =
O(\log(1/\epsilon))$,
such that $\Pr[S_n < t - c | S_n < t] > \delta$
whenever $\Pr[S_n < t] > \epsilon$.
\end{lemma}
\begin{proof}
Use Lemma \ref{lemma-gap} to group the $X_i$ together into partial
sums $Y_i = \sum_{j=i n_0+1}^{i n_0 + n_0} X_j$ with the property that
for some $t$
$\Pr[Y_i < t] < \frac{1}{2}$ but $\Pr[Y_i < t - c] = \delta_0 > 0$.
(Note that $n_0$ does not depend on $\epsilon$, so it disappears into
the constant factor.)
Then apply Lemma \ref{lemma-gap-relative} to sums of these $Y_i$
variables to get the full result.
\end{proof}

\subsection{When the Race Ends}

In this section, we show that a race between $n$ independent delayed
renewal processes with bounded added delays ends in $O(\log n)$
rounds with at least constant probability.  In the
following section, we
translate this result, which appears as Corollary
\ref{corollary-race}, back into terms of the \algname{} algorithm 
to get Theorem \ref{theorem-noisy}.

\begin{theorem}
\label{theorem-race}
Let $\{X_{ij}\}$, where $i, j \ge 1$,
be a two-dimensional array of finite non-negative
independent identically
distributed random variables with a common distribution function $F$
that is not concentrated on a point.  
Let $\{\Delta_{ij}\}$, where $i \ge 1, j \ge 0$, be a
two-dimensional array of constants with $0 \le \Delta_{ij} \le M$ 
when $j \ge 1$.
Let $\{H_{ij}\}$, where $i, j \ge 1$, be a two-dimensional array of
independent random variables, each of which is equal to $\infty$ with
probability $h(n)$ and $0$ otherwise.
Define 
\[S'_{ir} = \Delta_{i0} + \sum_{j=1}^{r} \left(\Delta_{ij} +
X_{ij} + H_{ij}\right).\]
Assume that for any finite $t$, integer $r$, and $i\ne j$, 
$\Pr[S'_{ir} = S'_{jr} = t] = 0$.
Let $c$ be any integer constant greater than $0$.

Then there exists a constant $\delta > 0$, such that 
for any $n$, there exists $r=O(\log n)$ and $t$,
such that
\begin{displaymath}
\Pr
\left[
\forall i \> S'_{ir} = \infty
\vee
\left(
\exists i \le n: S'_{i,r+c} < t \wedge \forall i' \ne i, i' \le n: S'_{i'r} > t
\right)
\right]
>
\delta.
\end{displaymath}

The constant factor in $r=O(\log n)$ and
the constant $\delta$ may depend on $c$, $F$, $M$, and $h$; but
neither constant
depends on $n$.
\end{theorem}
\begin{proof}
Since each $X_{ij}$ is finite with probability $1$, there exists some
constant $c_1$ such that 
$\Pr[\sum_{j=r+1}^{r+c} X_{ij} < c_1] > \frac{1}{2}$.
Let $T_{ir} = \sum_{j=1}^{r} X_{ij}$
and let $S_{ir} = T_{ir} + \sum_{j=0}^{r} \Delta_{ir}$.
Apply Lemma \ref{lemma-gap-combined} to the sequence
$X_{ij}$ with $c = cM + c_1$
and $\epsilon = n^{-2}$ 
to obtain $r = O(\log n)$ and a constant $\delta_0$ for which
$\Pr[T_{ir} < t-cM - c_1 | T_{ir} < t] > \delta_0$
whenever
$\Pr[T_{ir} < t] > n^{-2}$.
Adding the missing constant terms $\sum_{j=0}^{r} \Delta_{ij}$ 
to $T_{ir}$ to get $S_{ir}$
is equivalent to subtracting these same terms from each occurrence of
$t$,
so we in fact have
$\Pr[S_{ir} < t-cM - c_1 | S_{ir} < t] > \delta_0$
whenever
$\Pr[S_{ir} < t] > n^{-2}$.
This gives us our target round $r$.

Now apply Lemma \ref{lemma-winner} 
to $S'_{ir}$, for all $i \le n$, to show that with probability at least $1/5$
either $\forall i S'_{ir} = \infty$ or
there exists a time $t_0$,
such that there is a unique winner $i \le n$ 
for which $S'_{ir}$ is less than $t_0$.
Let us assume without loss of generality that $n$ is at least
$6$.
Throw out all cases where $i$ has $\Pr[S'_{ir} < t_0] \le n^{-2}$;
this leaves a probability of at least $1/5 - 1/n \ge 1/30$ that
(a) there is a unique winner $i$, and (b) $i$ satisfies the condition
$\Pr[S'_{ir} < t_0] > n^{-2}$, implying
$\Pr[S_{ir} = S'_{ir} < t_0] > n^{-2}$ and thus
$\Pr[S_{ir} < t_0-cM - c_1 | S_{ir} < t_0] > \delta_0$.
So with probability at least $\frac{1}{30}\delta_0$,
we have $S_{ir} < t_0 - cM - c_1$,
and thus
with probability at least $\frac{1}{60}\delta_0$
we have
$S_{i,r+c} < S_{ir} + cM + c_1 = S'_{ir} + cM + c_1 < t_0$.

Suppose that this event holds.  It is still possible for $S'_{i,r+c}$
to be infinite if $\sum_{j=r+1}^{r+c} H_{ij} = \infty$.  Call this
event $I$; if $\Pr[I] = 1 - (1-h(n))^c > \frac{1}{120}\delta_0$, then
$h(n)$ is bounded below by a constant and there exists $r' = O(\log
n)$ such that $\Pr[\forall i S'_{ir'} = \infty]$ is at least a
constant.  Alternatively, we have 
\(
\Pr[S'_{i,r+c} = S_{i,r+c} < S'_{ir} + cM + c_1] 
> \delta = \frac{1}{120}\delta_0.
\)
In either case, the theorem holds.
\end{proof}

\begin{corollary} 
\label{corollary-race}
Let 
$R$ be the first round for which either
\begin{itemize}
\item
There
exists $i$, such that 
$S'_{i,R+c} < S'_{i'R}$ for all $i' \ne i$, or
\item
For all $i$, $S'_{i,R+c} = \infty$.
\end{itemize}
Under the conditions of the preceding theorem, 
$\E[R] = O(\log n)$, and, for any $k \ge 0$,
$\Pr[R > k] \le e^{-\floor{k/O(\log n)}}$.
\end{corollary}
\begin{proof}
Theorem \ref{theorem-race} says that the
desired event occurs with constant probability $\delta$ after 
a phase consisting of $r = O(\log n)$
rounds.  If it does not occur, we can apply the theorem again to the
subset of the $i$'s for which $S'_{i,r+c}$ is finite,
starting with round $r+c+1$ and setting
the initial delay $\Delta_{i0}$ to the value of
$S'_{i,r+c}$ from the previous phase.

On average, at most $1/\delta = O(1)$ such phases are needed, giving
$\E[R] \le (1/\delta) r = O(\log n)$.
For the exponential tail bound, observe that the probability that the
algorithm runs for more than $c$ phases of $r$ rounds each is at most
$(1-\delta)^{c} = \left((1-\delta)^{1/\delta}\right)^{c \delta}
\le \left(e^{-1}\right)^{c \delta} = e^{-c \delta}$.
So the probability that the algorithm runs for more than $k$ rounds is
at most $e^{-\floor{k/r}\delta} \le e^{-\floor{k/O(\log n)}}$.
\end{proof}

\subsection{When \algname{} Ends}

Translating Corollary \ref{corollary-race} back into terms of the
\algname{} algorithm gives:

\begin{theorem}
\label{theorem-noisy}
Under the noisy scheduling model with random failures,
starting from any reachable state in the \algname{} algorithm in which
the largest round number of any process is $r$, the algorithm running
with $n$ active processes
terminates by round $r+r'$, where $r'$ has expected value $O(\log n)$
and $\Pr[r' > k] \le e^{-\floor{k/O(\log n)}}$ for any $k \ge 0$.
\end{theorem}
\begin{proof}
Apply Corollary \ref{corollary-race} with $c=2$ and the initial delays
$\Delta_{i0}$ set to the times at which each process completes round
$r$.  This shows that after $R$ additional 
rounds, where $\E[R] = O(\log n)$ and
$\Pr[R > k] \le e^{-\floor{k/O(\log n)}}$, either some process
$P$
finishes some round $s$ before any other process finishes round $s-2$,
or all processes fail.
In the first case, if $P$ prefers $b$, it is the only process
to have written to $a_b[s-1]$ or $a_{1-b}[s-1]$ by the time it reads
$a_{1-b}[s-1]$ as part of round $s$.  
Thus it reads a zero from $a_{1-b}[s-1]$ and decides.  All other
processes decide at most one round later by Lemma
\ref{lemma-agreement}.  We thus get $r' \le R+1$, and the single extra round
disappears into the constant factors.
\end{proof}

It is not hard to see that an $O(\log n)$ bound
is the best possible, up to
constant factors.

\begin{theorem}
\label{theorem-lower-bound}
There exists a noise distribution $F$ and a set of delays $\Delta$
such that the \algname{} algorithm requires expected $\Omega(\log n)$
rounds in the noisy scheduling model, even without failures.
\end{theorem}
\begin{proof}
Let all $\Delta_{ij} = 0$ for $j > 0$, and let $F$ have each
operation take either $1$ or $2$ time units with equal probability.
Then any single processor completes its first $\log n$
operations in $1$ time unit each with probability $1/n$.
To avoid simultaneous operations, let $\Delta_{i0}$ be some small
distinct epsilon value for each $i$.

Start $n/2$ processes with input $0$ and $n/2$ with input $1$.
The probability that there exists at least one $0$-input process and
at least one $1$-input process that both complete their first $\log n$
operations in $1$ time unit each is given by
\begin{displaymath}
\left(1-\left(1-\frac{1}{n}\right)^{n/2}\right)^2
\end{displaymath}
which goes to $(1-e^{-1/2})^2 = \Theta(1)$ in the limit as $n$ grows.
So there is a constant probability that at least one
process with each input runs for $\log n$ operations without ever
changing its preference to that of
a faster process with the opposite preference, and we get expected
$\Omega(\log n)$ rounds of disagreement.
\end{proof}

\section{Termination with Quantum and Priority-Based Scheduling}
\label{section-hybrid-termination}

In this section, we consider the question of termination subject to
hybrid quantum and priority-based scheduling on a uniprocessor.  The
required quantum size is 8 operations; curiously, this is the same
size required for the specialized algorithm given in
\cite{AndersonM1999}.  We see this coincidence as hinting at the
possibility 
that all shared-memory
consensus algorithms may ultimately converge to a single ideal
algorithm (though such an ideal algorithm, if it exists, 
is probably not identical to \algname{}).

\begin{theorem}
When running \algname{}
in a hybrid-scheduled system with
a quantum of at least 8 operations, 
every process decides after executing at
most 12 operations.
\end{theorem}
\begin{proof}
We will show that at most one of $a_0[1]$ and $a_1[1]$ is set before
some process finishes round $2$ and decides.
Consider an execution in which $a_0[1]$ and $a_1[1]$ are each set at
some point.  Let $P_0$ and $P_1$ be the first processes to set
$a_0[1]$ and $a_1[1]$, respectively.  Neither $P_0$ nor $P_1$ can have
observed the round-$1$ write of the other, or it would have changed
its preference.  Thus both processes' round-$1$ reads of $a_0[1]$ and
$a_1[1]$ must have occurred before either performed its round-$1$
write.  Since we are on a uniprocessor, this can only occur if one of
the processes was pre-empted before its write occurred.

Assume without loss of generality that $P_0$ is this unlucky process.
Since $P_0$ is the first process to write to $a_0[1]$, if we can show
that $P_0$ is not rescheduled before some process completes round $2$,
then that process decides $1$ (and by Lemma \ref{lemma-agreement}, all
processes eventually decide $1$) as soon as it observes a zero in
$a_0[1]$.  So we need only show that $P_0$ is not rescheduled until
some other process completes eight operations.

Let $Q_1$ be the process that pre-empts $P_0$.  At the time of
pre-emption, $Q_1$ is at the start of a quantum; it either finishes
eight operations without being pre-empted or is pre-empted by a
higher-priority process $Q_2$.  But $Q_2$ in turn can only be
pre-empted before completing its quantum by some higher-priority
process $Q_3$.  After at most $n$ such pre-emptions, we run out of
higher-priority processes, and the last process runs to the end of its
quantum and decides.  Note that all of the processes in this chain
(except possibly $Q_1$) have a higher priority than $P_0$ and thus
cannot be equal to $P_0$.  It follows that some process finishes round
$2$ before $P_0$ is rescheduled, and thus every process decides $1$ by
the end of round $3$.
\end{proof}

\section{Bounded space \algname}
\label{section-bounded-memory}

The \algname algorithm as described in Section~\ref{section-algorithm}
requires infinite space.  In this section, we describe how to modify
the algorithm to use bounded space.  We assume that we have available
a \emph{backup protocol}, which is a
bounded-space consensus protocol that requires polynomial work per
process (for example, the $O(n^4)$ protocol in \cite{Aspnes93} works).
We will build a protocol that combines \algname with the backup
protocol in a way that only uses the backup protocol rarely, so that
its high cost adds only a constant to the $O(\log n)$ cost of the
combined protocol.

Note that such a combined protocol is not necessary in the model of
Section~\ref{section-hybrid-termination}, as in that model we only
need space for $3$ rounds of \algname.

The combined protocol operates as follows:
\begin{enumerate}
\item Run \algname through round $\rmax$.
\item At round $\rmax+1$, switch to the backup protocol, using the
preference at the end of round $\rmax$ of \algname as input to the backup
protocol.
\end{enumerate}

If $\rmax$ is large enough, most of the time we will expect that
\algname{} terminates before reaching $\rmax$ and the backup
algorithm will not be used.  But in the case where $\rmax$ is
reached (say, because the scheduler is nastier than we have assumed),
the backup algorithm guarantees termination using bounded space and
bounded (but possibly very large) expected time.  

\begin{theorem}
\label{theorem-bounded-memory}
For any polynomial-work consensus protocol chosen as a backup
algorithm and any noise distribution, 
there is a choice of $\rmax = O(\log^2 n)$ such that the combined
algorithm described above is a consensus protocol that
requires $O(\log n)$ expected operations per
process and $O(\log^2 n)$ bits in the $a_0$ and $a_1$ arrays.
\end{theorem}
\begin{proof}
First let us show that the combined algorithm solves consensus.
Validity is immediate from Lemma~\ref{lemma-validity}; when all inputs
are equal, we never get past round $2$ and the combined algorithm
behaves identically to \algname.
For agreement, the only tricky case is when some processes decide
during \algname and others decide during the backup protocol.  But if
some process $P$ decides $b$ at or before round $r$, then by
Lemmas~\ref{lemma-converging-preferences} and~\ref{lemma-agreement}
no process writes $a_{1-b}[r]$ and every process that executes the
backup protocol has $b$ as input.  Thus the validity condition for the
backup protocol implies that all processes decide $b$.

Now let us show that there is a choice of $\rmax$ that gives the
desired performance bound.  Suppose each process finishes the backup
protocol in $O(n^c)$ expected operations.  By
Theorem~\ref{theorem-noisy}, there is a value $T = O(\log n)$ such
that the probability that \algname does not finish by round $k$ is at
most $e^{-\floor{k/T}}$.  
Setting $\rmax = T\cdot c\cdot \log n =
O(\log^2 n)$,
the backup protocol is run with probability at most $e^{-c \log n} =
n^{-c}$, and thus it contributes at most $n^{-c} O(n^c) = O(1)$ to the
expected cost.

Finally, the size of the $a_0$ and $a_1$ arrays is clearly equal to
$\rmax = O(\log^2 n)$.
\end{proof}

\section{Simulation Results}
\label{section-simulations}

\begin{figure*}
\begin{center}
\input{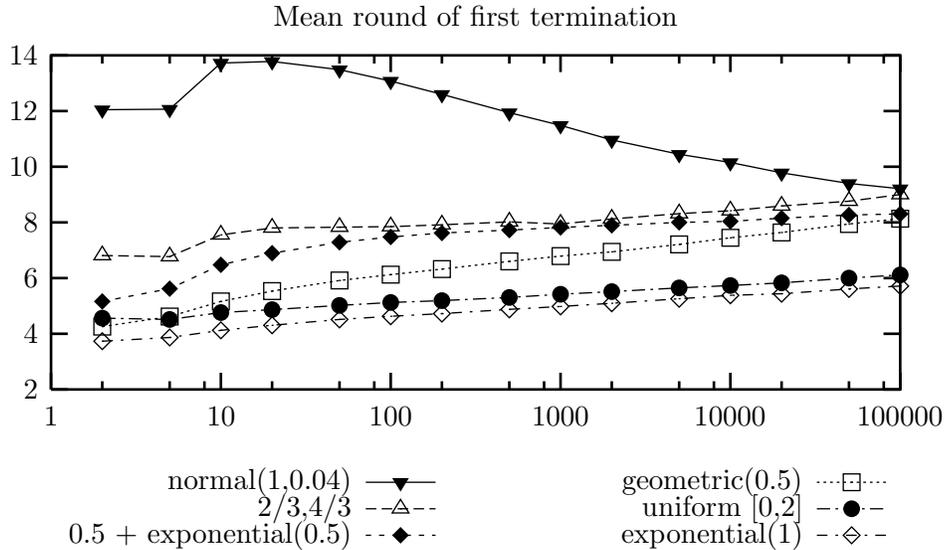}
\end{center}
\caption{Results of simulating \algname{} with various interarrival
distributions.}
\label{figure-rounds}
\end{figure*}

Figure \ref{figure-rounds} gives the results of simulating \algname{}
with various interarrival distributions.  These simulations are of the
model as described in Section \ref{section-noisy}; in
particular it is assumed that all operations take zero time and that
there are no contention effects or synchronization issues.

The X axis is plotted on a logarithmic scale and represents the number
of processes.
The Y axis is plotted on a linear scale and represents the round at
which the first process terminates (which may be one less than the round
at which the last process terminates).
Each point in the graph represents an average termination round in
10,000 trials with the given distribution and number of processes.
The starting times for all processes are the same except for a small
random epsilon, generated uniformly in the range $(0,10^{-8})$.
In each case, half the processes are started 
with input 0 and half with
input 1.
There are no failures.

The random number generator used was {\tt drand48}. 
The distributions used were:
\begin{enumerate}
\item Normal distribution with mean 1 and standard deviation 0.2
(variance 0.04), 
rejecting points outside $(0,2)$.
\item $2/3$ or $4/3$ with equal probability.
\item $0.5$ plus an exponential random variable with mean $0.5$.  This
corresponds to a delayed Poisson process.
\item Geometric with $p = 0.5$.
\item Uniform in $(0,2)$.
\item Exponential with mean $1$.  This corresponds to a Poisson process
with no initial delay; it is also equivalent to generating a schedule
by choosing one process uniformly at random for each time unit.
\end{enumerate}

It is worth noting that while the expected number of rounds grows
logarithmically for most distributions, both the rate of growth and
the initial value are small.
These small constant factors may be the result of most processes adopting
the values of early leaders, so that termination can be reached by
agreement among leaders rather than the emergence of a single leader.

The inverted behavior with a normal
distribution is intriguing; it suggests that with large numbers of
processes there are more chances for one particularly speedy process
to leap ahead of its competitors, and that for some distributions this effect
overshadows the effect of having more competitors to leap ahead of.
It is not clear from the data
whether this curve eventually turns around and starts rising
again, or whether it converges to some constant asymptote.

\section{Conclusions, Extensions, and Future Work}
\label{section-extensions}

We see this paper as making two main contributions.  The first is the
extraction of the adaptive $\Theta(\log n)$ time 
\algname{} protocol from its more sophisticated
predecessors and the demonstration that this simplified algorithm can
solve consensus in models that are less extreme than those
predecessors were designed to survive but that are perhaps closer to
capturing the scheduling behavior an algorithm is likely to experience
in practice.  Although \algname{} does not really contain any new
ideas, we believe that ripping out features
that practitioners might
balk at implementing is a valuable task in its own right.

The second is the noisy scheduling model.  This model limits the
adversary not by covering its eyes but by making its hands shake.
It allows us to express the understanding that in the real world
failures and timing
are usually not fully under the control of intelligent
demons, while still retaining a healthy respect for the subtlety and
unpredictability of the world.
We believe that this ``perturbed worst-case analysis'' approach 
is likely to have applications in many areas both in and outside of
distributed computing.

There are still many questions left unanswered and many ways in
which the noisy scheduling model could be extended.  We
discuss some of these issues below.

\paragraph*{Non-random failures.}
It would be nice to understand how \algname{} fares with failures that
are not random.
We can get an upper bound in this situation
by restarting Theorem \ref{theorem-noisy} whenever a process dies.
Since the adversary must kill at least one process every expected $O(\log n)$
rounds, the algorithm terminates in expected $O(f \log n)$ rounds
where $f$ is the number of failures.  This bound compares favorably with the
$O(n \log^2 n)$ work per processor needed by the best known randomized
algorithm that solves consensus with a fully-adaptive adversary
and up to $n-1$ failures
\cite{AspnesW1996},
but the fully-adaptive adversary is 
much stronger than 
one limited to noisy scheduling.
It seems likely that a better upper bound than $O(f \log n)$ could be
obtained by a more careful analysis that includes how processes change
preferences.  We conjecture that the real bound is in fact $O(\log n)$.

\paragraph*{Statistical adversaries.}
We would also like to do away with
the fixed bound $M$ on the delay between operations under the control
of the adversary.  The technical reason for including this
bound in the model is that it provides a scale for the noise
introduced by the $X_{ij}$ variables; if the adversary can increase
$\Delta_{ij}$ without limit, it can construct a steadily slower and
slower execution in which the
noise, relative to the gap between rounds, never accumulates enough to
affect the schedule.  But a weaker statistical constraint, such as
requiring $\sum_{j=1}^{r} \Delta_{ij} \le rM$, might avoid such
Zeno-like pathologies while allowing more variation in the gaps
between operations.\footnote{This is a bit like using the
\emph{statistical adversary} of
\cite{ChouCEKL95}.} 
The present proof does not work with just this
statistical constraint (the particular step that breaks down is
the use of Lemma
\ref{lemma-gap-combined} to show that being ahead at round $r$ often
means being ahead by $c$ at round $r$), but we conjecture that 
the statistical constraint is in fact enough 
to get
termination in $O(\log n)$ rounds.

\paragraph*{Synchronization and contention.}
Though the present work was motivated by a desire to move away from
powerful theoretical adversaries toward a model more closely
reflecting the non-maliciousness of misbehavior in real systems, we
cannot claim that the model accurately describes the behavior of any
real shared-memory system.  One difficulty is that real shared-memory
systems generally
do not guarantee full serializability of memory operations in
the absence of additional synchronization operations
(see \cite[Section 8.6]{PattersonHG1996}).
We can overcome this difficulty by adding
synchronization barriers to each round of \algname{}; in principle
this does not affect the analysis since the structure of each round is
still the same as all other rounds.  
A second problem is memory contention, which we have not analyzed.
The difficulty with both explicit synchronization and memory
contention is that their effects are unlikely to be consistent with
the assumption that the timing of different processes' operations are
independent.  To the extent that this lack of independence disperses
processes (say, by slowing down laggards fighting over congested early-round
registers while allowing the speedy to sail through relatively
clear late-round registers), it helps the algorithm.  Whether such
an effect would occur in practice cannot easily be predicted without
experimentation.

\paragraph*{Lower bounds.} The noisy scheduling model is friendly
enough that an $O(\log n)$ running time for consensus might not be
the best possible.  A counterexample like the one given in the proof of
Theorem \ref{theorem-lower-bound} might be able to show that no
deterministic algorithm with certain strong symmetry properties (such
as no dependence on process identity and a mirror-image handling of
the different inputs) can do better, 
but it not obvious where to look for a more general lower bound.
It is not out of the question that a clever algorithm could solve
consensus with noisy scheduling in as little as $O(1)$ time.

\paragraph*{Message passing.}
All of our results are set in a shared-memory model.  It would
be interesting to see whether a noisy scheduling assumption can be
used to solve consensus quickly in an asynchronous message-passing
model.

\paragraph*{Other problems.}
Finally, though we have concentrated on a particularly simplified
protocol for solving a single fundamental problem, it would be
interesting to see how other algorithms fare in the
noisy scheduling model.  It seems likely, for example, that
algorithms designed for unknown-delay models such as Alur et al.'s
\cite{AlurAT1997} should continue to work in the noisy scheduling
model, perhaps with some constraint on the noise distribution to
exclude random delays with unbounded expectations.  Similarly the line
of inquiry started by Gafni and Mitzenmacher \cite{GafniM1999}, on analyzing
the behavior of timing-based algorithms 
for mutual exclusion and related problems
with random scheduling,
could naturally extend to the more general model of noisy scheduling.

\section{Acknowledgments}

I would like to thank Faith Fich and Maurice Herlihy for
insightful comments on the plausibility of an early version of the
noisy scheduling
model; 
the remaining implausibility
is my fault and not theirs.  
I am also
indebted to Robbert van Renesse for pointing out the ``narrowness'' of
the bad execution paths needed to prevent consensus as a reason for the
relative lack of concern for asynchronous impossibility results
among practitioners.

\bibliographystyle{plain}

\bibliography{race}

\end{document}